\documentclass[preprint,aps]{revtex4-1}
\usepackage{graphicx}
\usepackage{dcolumn}
\usepackage{bm}
\usepackage{amsmath}
\bibstyle{biophysj.bst}
\usepackage{graphicx,pstricks,color} 

\usepackage{natbib}
\setcitestyle{square, comma, numbers, sort&compress, super}
\usepackage{upgreek}
\usepackage{enumerate}

\usepackage[mathlines]{lineno}
\usepackage{listings}

\begin{document}
\title {Strain softening and stiffening responses of spider silk fibers probed using Micro-Extension Rheometer
}

\author{Sushil Dubey$^\dagger$, Chinmay Hemant Joshi$^{\$\S}$, Sukh Veer$^{\dagger\S}$, Divya Uma$^{\#}$, Hema Somanathan$^\$$, Sayantan Majumdar$^\dagger$, and Pramod A Pullarkat$^{\dagger}$}
\email{Corresponding author: pramod@rri.res.in; $^{\S}$ These authors contributed equally.} 
\affiliation{
$\dagger$ Soft Condensed Matter Group, Raman Research Institute, C. V. Raman Avenue, Bengaluru, Karnataka 560 080, India\\
$\$$ IISER TVM Centre for Research and Education in Ecology and Evolution (ICREEE), School of Biology, Indian Institute of Science Education and Research, Thiruvananthapuram, Kerala 695 551, India\\
$^{\#}$ School of Liberal Studies, Azim Premji University, Bengaluru, Karnataka 560 100, India
}

\begin{abstract}
Spider silk possesses unique mechanical properties like large extensibility, high tensile strength, super-contractility, etc. Understanding these mechanical responses 
require characterization of the rheological properties of silk beyond the simple force-extension relations which are widely reported. Here we study the linear and non-linear viscoelastic 
properties of dragline silk obtained from social spiders {\it Stegodyphus sarasinorum} using a Micro-Extension Rheometer that we have developed. Unlike continuous extension data, our technique 
allows for the probing of the viscoelastic response by applying small perturbations about sequentially increasing steady-state strain values. In addition, we extend our analysis to 
obtain the characteristic stress relaxation times and the frequency responses of the viscous and elastic moduli. Using these methods, we show that in a small strain regime (0--4\%) dragline
silk of social spiders shows strain-softening response followed by strain-stiffening response at higher strains ($> 4\%$). The stress relaxation time, on the other hand, increases monotonically with increasing strain for the entire range. We also show that silk stiffens while ageing within the typical lifetime of a web. Our results demand the inclusion of the kinetics of domain unfolding and refolding in the existing models to account for the relaxation time behaviour.
\end{abstract}

\maketitle

\section*{Introduction}

Spider silk is one the strongest bio-material known to mankind  and unlike man-made
fibers, silk exhibit unique mechanical properties like large extensibility, supercontractility, etc. \cite{harmer2011, vollrath2000, yarger2018, Blackledge2009, liu2005, book-foelix2011}.  These properties 
have been fine tuned by evolution to match the specific requirements demanded by the type of web that the spider constructs, the strains the web is likely to experience, and its life 
cycle. Spiders produce silk and store in the silk gland in the form of a dope  and spin fibers using the dope whenever required via specialised orifices called as spinnerets \cite{book-foelix2011, gosline1986}.
Spiders are capable of producing up to seven different silk types, each of which is engineered for a particular function through millions of years of evolution \cite{harmer2011, vollrath2000, book-foelix2011}. Dragline silk is produced on a large scale by all spiders, as it forms the main structural part of any spider web and is also used by spiders to escape from predators \cite{harmer2011, vollrath2000, book-foelix2011, gosline1986}. Dragline silk is the most easily procurable and widely studied among all other spider silk types \cite{harmer2011, vollrath2000, yarger2018}. There are a plethora of studies done on mechanical properties of spider silk under various conditions like differential humidity and temperature \cite{plaza2006, vehoff2007}, exposure to various solvents \cite{liu2005, shao1999, vollrath2001}, different reeling speeds \cite{liu2005, vollrath2001}, 
manipulation of nutrition of spiders \cite{madsen1999}, etc. Capture silk is another widely studied form of spider silk. There are two major types of capture silk: cribellate silk and viscid silk \cite{harmer2011}, and there are some studies done on both capture silk types \cite{Blackledge2006, BLACKLEDGE2005, Eberhard1993, Opell2001, Sahni2011}. Other silk types like Acniform, Tubiliform, and Piriform are less studied owing to lack of their abundance and difficulty in procurement of good quality samples \cite{Blackledge2006b, Jiang2011}.

Social spiders species {\it Stegodyphus sarasinorum} are of advantage, as they exhibit nest-site fidelity and thus can easily be maintained under controlled laboratory conditions (light, humidity, temperature, etc.) which could affect silk fibers mechanical properties. One of the novelties of our study is that so far only silk of solitary spider species have been investigated. Social spider colonies can consist of over a hundred individuals and their webs can be very large - spreading over a meter under field conditions Fig. S1(A). This makes it interesting as a model to study the physical properties of their webs.  They are also re-used by subsequent generations with continuous repair and maintenance \cite{Lubin2007, Avils1997}.

Investigations of silk mechanics are typically performed by generating large force-extension loops by applying cyclic linear stretching or single stretch up to the yield or break point. Various quantities like yield stress, yield strain, breaking stress, breaking strain, Young's modulus and dissipation are calculated using this data \cite{yarger2018, vehoff2007, shao1999, Swanson2006}. However, a full characterisation of the linear and non-linear rheological properties of single strands of silk require the ability to apply a variety of rheological protocols in extension mode, which are limited by available techniques. To overcome this, we 
have investigated dragline
silk from social spiders using an improved version of the Micro-Extension Rheometer (MER), which we had developed earlier \cite{giri2013}. In order to obtain the elastic moduli about equilibrium states with different amounts of pre-strain, we used a sequential step strain protocol where the silk is pre-strained by increasing amounts, and the modulus is measured at steady-state using small amplitude strain oscillations. We observe that silk exhibits an initial strain-softening for up to about 4\% strain and then begin to stiffen with further increase in strain. An important feature of our study is that the stress relaxation corresponding to each applied strain step is used to obtain the relaxation times as a function of strain. Remarkably, the relaxation times thus obtained increase with 
increasing strain and tend to saturate at high strains. Using a Fourier analysis method, we also extract the frequency dependence of the elastic and viscous moduli from the relaxation process. The rheological 
data presented here, especially the frequency response and the relaxation time vs. strain data will aid in relating molecular level processes to the unique macroscopic viscoelastic properties of silk. Another important feature of our study is that we preserve the natural pre-tension of the silk fiber in the web while transferring it for rheological investigations. This pre-tension can be obtained from the sequential step strain data.  

\section*{Material and Methods:}

\subsection*{Preparation of silk samples}

\subsection*{(i) Colony setup} Three spider colonies were collected from Somapura, Karnataka and Kuppam, Andhra Pradesh, India. 
A group of twenty spiders was randomly selected from the colonies. Each such group was introduced on a square metal wire frame 
(45cm x 45cm; wire dia. 0.1cm) Fig. S1(B). Social spiders construct sheet webs on the surface of which prey are intercepted and the sheet web is anchored to plants using dragline silk. The spiders live within nests made of dense silk (also called retreats) and come out to hunt when they sense vibrations from an entangled prey \cite{Lubin2007, Avils1997}.
To mimic this, a small amount of the retreat material was placed between two cardboard sheets (6cm x 6cm) attached to the frame such that there is a gap between the two sheets.
Spiders were introduced into this gap where they reside when inactive. Usage of a thin frame enabled spiders to build a two-dimensional web (Fig. S1(B)) and made it easy to 
extract the silk. The frames were suspended on a stand and a thick layer of petroleum jelly was applied at the point of suspension to avoid ant attack and to prevent spiders from escaping.
The whole setup was kept in a room with enough ventilation and with 12h:12h day-night cycle. Spiders were kept undisturbed for three days, before the silk extraction process.
During the whole experiment, a little water was sprayed on the spiders twice a week using a water sprayer and were fed two grasshoppers in a week.

\subsection*{(ii) Silk extraction process}
Dragline silk is the main structural silk and cribellate silk is the capture silk in {\it S. sarasinorum} webs (Fig. S1(C)) \cite{Bradoo1972}. In order to obtain fresh silk samples every day, a small area of the
spider web was disturbed one day prior to silk collection. Spiders repaired that disturbed area overnight by spinning fresh silk strands of both silk types. For
extraction, a Y-shaped cardboard frame (width: 1.5 cm) (Fig. S1(D)) with a small speck of super-glue
(Fevi Kwik\textsubscript \textregistered ) on both the arms of the ``Y" was stuck to the silk fiber on the web. After a few seconds, the surrounding fiber was cut using  a pair of scissors. This process ensured
that the pre-tension of the silk is not lost. After this, the Y-frame (with the silk fiber on it)
was embedded on a thermocol sheet and was stored in a plastic box, before performing tensile measurements. Except for measuring age effects,
all samples were studied within a day of the extraction.

\subsection*{(iii) Sample preparation} 
For performing force measurements, we prepared a sample mount consisting of two microscope coverslips placed side by side and parallel to each other 
with a gap of 300-500 $\mu$m. The two coverslips were held together by a glass slide to which they were glued (Fig. S1(E)).
Silk fiber on the Y-frame was placed over the coverslips such that the fiber forms a bridge between the two coverslips, and was glued close to the edges of 
the coverslips. The sample was kept for
drying at room temperature (24--26 $^\circ$C; rel. humidity 50--60\%). The sample was then transferred on to the microscope stage for rheological measurements.
\vspace*{-1cm}
\subsection*{Rheological Measurements ($\lesssim 4$\% strain)} 
Measurements were done using a modified version of the home-developed Micro-Extension Rheometer (MER) which is an optical fiber based force device, described in 
detail in Ref. \cite{giri2013} (Fig. S1(F), S2, S3). For this, cylindrical optical fiber glass cantilevers
(125 $\mu$m in diameter and lengths of 2.5 mm or 5 mm) were made by cutting a single mode optical fiber  (P1-630A-FC, Thorlabs Inc., United States) with a scalpel. The base of the cantilever was attached to a linear piezoelectric drive (P-841.60, Physik Instruments GmbH, Germany) which has a position accuracy of 1 nm and a travel range of 94 $\mu$m. The piezo was mounted on an inverted microscope (Zeiss Observer.D1, Carl Zeiss GmbH, Germany)  using a motorized XYZ stage (XenoWorks, Sutter Instruments, USA). 
Laser light (HeNe 632.8 nm) exiting the tip of the cantilever was focussed on to a Position Sensitive Detector (PSD) (S2044, Hamamatsu Photonics, Japan) mounted on a side imaging port of the microscope. The cantilever deflection was calculated as the difference between the applied piezo displacement and the measured cantilever tip displacement with a final resolution of 35 nm and 70 nm for 40$\times$ and 20$\times$ objectives, respectively. The piezo and the detector were interfaced via a PC to enable automation and operation in constant strain mode using a feedback loop implemented using a home developed LabView code.  The green microscope illumination light and the red laser light were separated using appropriate filters to enable simultaneous force measurements using PSD and imaging using a CCD camera (Andor Luca R604, Andor Technology, Ireland). A photograph of the setup is shown in Fig. S3. The silk fiber glued between the two parallel cover slips (Fig. S1(E)) was placed on the microscope stage and
was displaced in the transverse direction using the tip of the optical fiber placed at the mid-point of the silk strand (Fig. S1(F), Fig. S2). The desired strain was calculated from the geometry shown in Fig.~\ref{forcedg}(A)  as $\gamma=(\sqrt{L_{i}^{2}+4 d^{2}}-L_{i})/L_{i}$. The deflection of the cantilever is calculated as $\delta = D-d$ where $D$ is piezo displacement and $d$ is displacement of the tip of the 
cantilever obtained from the PSD output. The force on the cantilever is then given by $F = -k\delta$, where $k$ is the cantilever force constant. The force constant is obtained by
placing the cantilever in a horizontal position and loading it with small pieces of thin copper wires of known weight.


\subsection*{Large strain experiments ($\gtrsim 4$\% strain)}
The maximum strain that can be applied using the piezo drive is limited by the travel range of the piezo (94 $\mu$m). In order to probe the response of the silk fibers at higher strains, we manually pulled the silk fibers using the motorized XYZ stage holding the piezo. This stage is controlled by a
Joystick with selectable speeds. Once the stress relaxation is complete, the displacement of the optical fiber was calculated from the camera images. 
Then sinusoidal oscillations were applied using the piezo to measure the viscoelastic response at that value of strain. Successive manual displacements are applied in this manner 
 to obtain different pre-strain levels. With this method, we were able to apply up to 30\% pre-strain.

\subsection*{Microscope imaging} 
The silk fibers mounted on the microscope were imaged using 20$\times$ objective to measure its length, and a 40$\times$ objective to measure its diameter. 
The pixel size of the final CCD images were calibrated using a micrometer scale (Olympus Objective Micrometer). The fibers were also imaged during
experiments. Recovery of strain was also tested by releasing a few fibers from their maximally strained state. Typical radii values for silk is 1-2 $\mu$m.

\subsection*{Electron microscopy} 
Electron microscopy images were taken using Carl Zeiss Ultraplus-FESEM (EHT = 3-5 kV). Silk fibers were glued on ITO glass and then sputter-coated with platinum (coating thickness ~2 nm). Diameters for the silk 
strands were measured using Smart SEM software for EM images. Typical EM images of dragline and cribellate silk are shown in Fig. S4. 
        
\section*{Results}

\subsection*{Viscoelastic response at small strain regime}

\begin{figure}
         \begin{center}
         \includegraphics[width=5.5in]{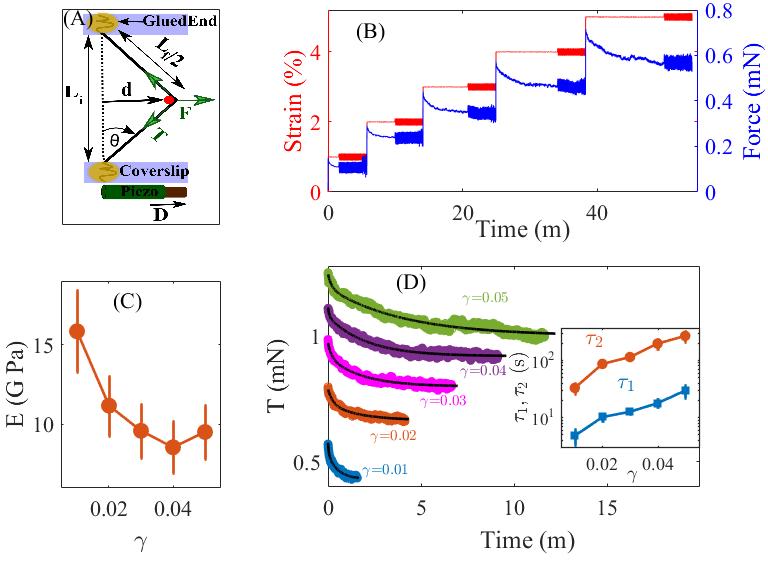}
         \caption{ (A) Schematic of a silk strand pulled using the optical fiber. The silk strand is stuck using glue at the two ends and pulled in the transverse direction with the tip of the optical fiber (red dot; also see Fig. S1(F)). Applied strain was calculated based on the geometry as described in the main text.  
         (B) Typical force response of a dragline silk subjected to sequential strain steps with a wait time between steps. The force relaxes to a steady-state value after each step 
         and this steady-state force increases with strain showing that  silk behaves as a viscoelastic solid.
         Small amplitude sinusoidal oscillations has been superposed
         towards the end of the relaxation process to obtain the viscoelastic moduli for each pre-strain. We first applied 20 cycles at 0.1 Hz followd by 50 cycles of 1 Hz.  (C) Storage moduli calculated using the 
         sinusoidal oscillations (0.1 Hz) exhibit a strain-softening behavior. The data points are averages of nine strands and the error bars are standard error of the mean (SE). (D) The tension relaxation curves obtained for each step is fitted to a double exponential to extract the relaxation times $\tau_1$ and $\tau_2$. The black curves are the fits (R-square is greater than 0.95 in all the cases). A log-linear plot of the variation of these relaxation times as a function of strain is 
         shown in the inset (averages for n = 10 strands).}
         \label{forcedg}  
         \end{center}
         \end{figure}

The mechanical properties of the silk strands were probed by displacing the strands in the transverse direction using the optical fiber as shown in 
Fig.~\ref{forcedg}(A), and detailed under Materials and Methods.
In order to obtain the viscoelastic responses of silk at different strain values we applied successive strains steps and recorded the force relaxation response as
shown in Fig.~\ref{forcedg}(B). Once the relaxation is close to the steady-state value, small sinusoidal strain oscillations of a fixed amplitude and frequency were superposed
on to the pre-strain. The force relaxation shows the viscoelastic nature of dragline silk. It also shows that silk behaves as solid at long time as indicated by the steady-state average force (Fig.~\ref{forcedg}(B)). 
From these data, the tension along the silk fiber is calculated as $T(t) = F(t)/(2 \sin\theta)$, where $\sin\theta = \dfrac{d}{\sqrt{d^{2}+(L_{i}/2)^{2}}}$, $\theta$ is the angle
with respect to the initial position, $d$ is the displacement of the tip of the cantilever which is in contact with the silk mid point, and $L_{i}$ is the initial length of the 
silk strand (see Fig.~\ref{forcedg}(A)). 
Using $T(t)$ thus obtained, we calculate the elastic moduli for different pre-strain values using the small amplitude oscillations as follows. Corresponding to the imposed strain oscillations 
$\gamma(t) = \gamma_{0} \sin(\omega t)$, we obtain stress oscillation as $\sigma(t)= \sigma_{0} \sin(\omega t + \phi)$, where $\sigma(t) = T(t)/A$, $A$ being the 
cross-sectional area of the silk fiber. 
The storage moduli for different strains are then obtained as $E'(\gamma) = \dfrac{\sigma_{0}} {\gamma_{0}} \cos(\phi)$, and the viscous moduli as 
$E''(\gamma) = \dfrac{\sigma_{0}} {\gamma_{0}} \sin(\phi)$. The storage moduli exhibit a  strain-softening behavior as shown in Fig.~\ref{forcedg}(C) with a minimum at around 4\% strain. 
The loss moduli could not be determined as the phase lag between the applied strain oscillations and the stress response was below the detectable range of the setup. 
The tension relaxation curves obtained from the application of sequential step-strains could be fitted well to
double exponential function $T(t) = a\, \exp(-t/\tau_{1}) + b\, \exp(-t/\tau_{2}) + c$, where $\tau_{1}$ and $\tau_{2}$ are the two relaxation times (see Fig.~\ref{forcedg}(D)). These relaxation times
increase with the applied strain as shown in the inset of Fig.~\ref{forcedg}(D). All fibers recover to the initial straight  configuration after releasing from maximally strained state within a few hours. After this recovery, the silk fibers quantitatively reproduce almost the same force response. We have also calculated the pre-tension on the silk fiber when present in the web by extrapolating the steady-state tension vs. strain data to zero strain value (Fig. S5). This pre-tension varies in the range of 0--9 mN (Fig. S5 (B)).

The frequency range we could probe using the current version of the Micro-Extension Rheometer was limited by the time constant of the feedback loop that maintains a prescribed
strain during the experiments. But the frequency information is contained in the tension relaxation data and this can be retrieved as discussed in the next section.

\subsection*{Frequency dependence of moduli}
\begin{figure*} 
\centerline{\includegraphics[width=.95\linewidth]{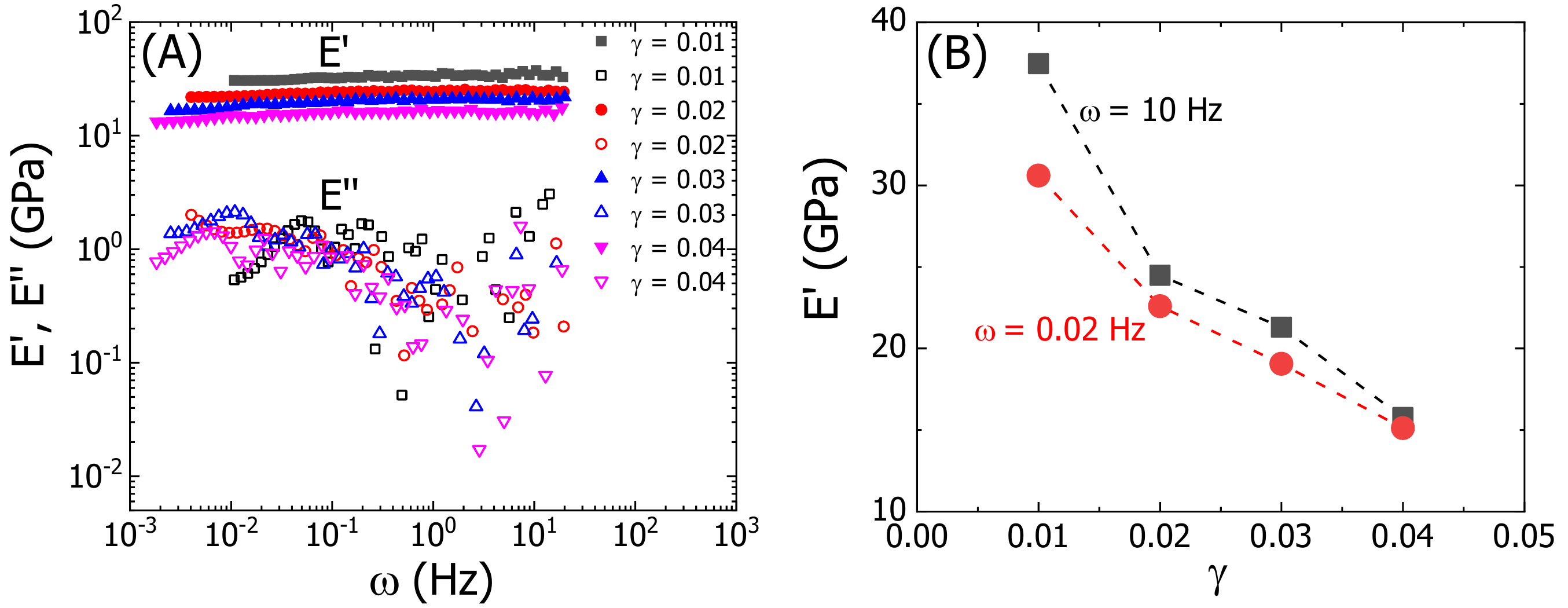}}
\caption{\label{FT}
  (A) Elastic moduli $E'$ (solid symbols) and viscous moduli $E''$ (hollow symbols) are plotted as a function of frequency for a dragline silk fiber for different step strain magnitude ($\gamma$ : 1 - 4 \%) indicated in the figure legend. (B) Variation of $E'$ with $\gamma$ for two different frequencies as indicated in the figure.   
}
\end{figure*}

Mechanical responses of viscoelastic materials are generally characterized by applying a sinusoidal stress or strain perturbation. The main advantage of such oscillatory measurement is that the steady-state response can be characterized independently as a function of the amplitude and frequency of the applied perturbations. To overcome the frequency limitation of
our experimental setup, we analyse the tension response to step-strain as follows. For a viscoelastic material, the linear response under an applied step-strain ($\gamma$) can be expressed in the time domain by stress relaxation modulus $E(t) = \frac{\sigma(t)}{\gamma}$, where $\sigma(t)$ is the time dependent stress in the system. Equivalently, such response can also be expressed in the frequency domain by a complex dynamic modulus $E^{*}(\omega) = E'(\omega) + i E''(\omega)$. Here, $E'$ is the storage modulus that quantifies the elastic response and $E''$, the loss modulus quantifies the viscous response of the material. In our experiments, we apply a strain deformation and measure the stress as a function of time. However, we observe that even for small strain amplitudes ($\sim$ 1\%), the accessible frequency range is below $\sim1$ Hz for the MER setup. In principle, the frequency dependent moduli can be obtained from time domain measurements, since $E(t)$ and $E^{*}(\omega)$ are related via simple Fourier transforms. However, in practice, reliable interconversion between $E(t)$ and $E^{*}(\omega)$ becomes nontrivial due to the noise present in the experimental data. A recent paper \cite{Evans2009} describes a simple and direct method to obtain frequency dependent moduli from time domain data that is not prone to the artefact due to noise. By using this method, $E^{*}(\omega)$ can be calculated from time dependent creep compliance $J(t) = \frac{\gamma(t)}{\sigma}$ obtained from step-stress experiments. Since, our experiments are strain controlled, direct measurement of $J(t)$ is not possible. Although, $E(t)$ and $J(t)$ are just reciprocal of each other dimensionally, they are mathematically related by a convolution integral
\cite{Ferry1980}: 
\begin{equation}
\int^{t}_{0}d\tau\, E(t - \tau)J(\tau) = t, \,\, (t > 0).
\label{E1}
\end{equation}  
If $E(t)$ or $J(t)$ varies smoothly and the variation is slow enough such that the magnitude of the local logarithmic slope, $n = \left|\frac{\partial \,log[\xi(t)]}{\partial \,log[t]} \right|$, where, $\xi(t)$ is either $E(t)$ or $J(t)$ remains small
($n \ll 1$), then $J(t) E(t) \approx$ 1 holds. The theory behind this approximation is well summarized in a recent paper \cite{Park1999}. 
We show the variation of $n$ as a function of time in Fig. S6. We find that for all values of strain, the maximum value of $n$, as estimated from the steepest portion of the curves, is $\ll$1. Thus, we can directly estimate $J(t)$ using the relation: $J(t) E(t) \approx$ 1. We show the variation of $E'$ and $E''$ of a dragline silk fiber as a function of $\omega$ obtained by this method 
for different step strain magnitudes in Fig.~\ref{FT}(A). We find that over almost 4 orders of magnitude variation in $\omega$, the fiber remains highly elastic with ($E' \gg E''$). The magnitudes of the elastic moduli compare well with the steady-state measurements presented in Fig.~\ref{forcedg}(C). Also, the flatness of $E'$ over wide frequency variation indicates the silk fiber is a viscoelastic solid over the observed frequency range. The gradual drop in $E'$ with increasing magnitude of applied step strains is consistent with the softening seen with increasing pre-strains (Fig.~\ref{forcedg}(C)). Such drop in $E'$ as a function of strain is shown for two values of frequency in Fig.~\ref{FT}(B). 
      
\subsection*{Viscoelastic response at large strain regime}

\begin{figure}[]
           \begin{center}
         \includegraphics[width=6in]{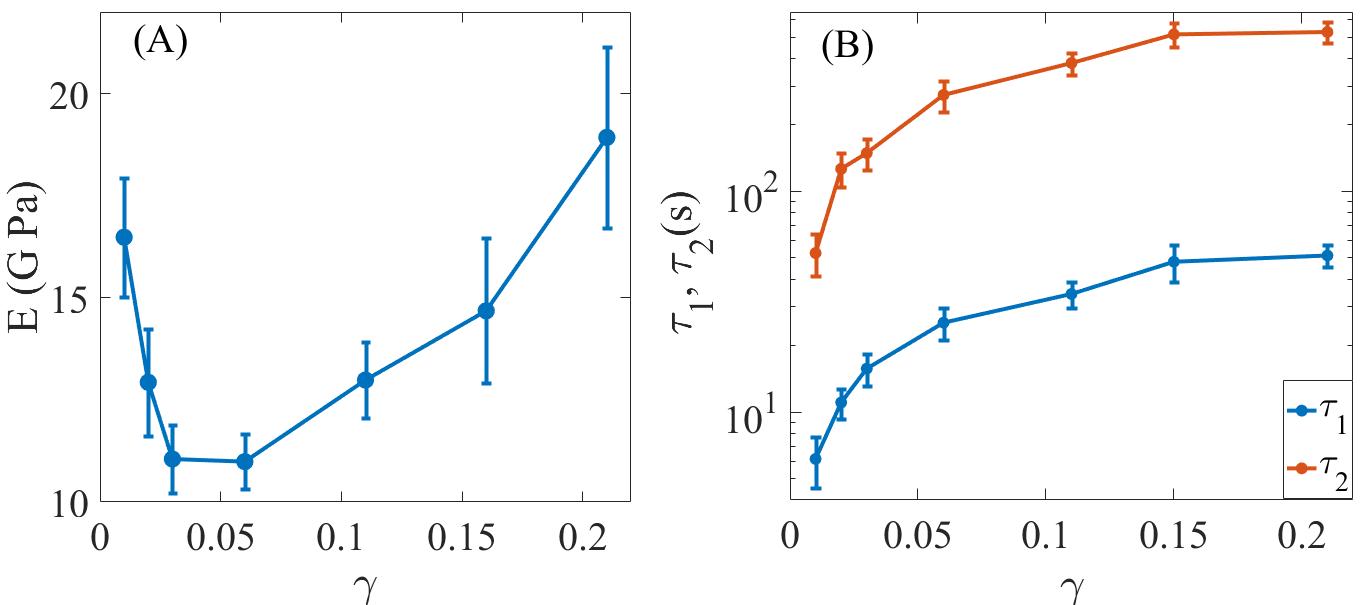}
          \caption{(A) Averaged elastic moduli obtained from large-strain experiments (n=10) show an initial strain-softening followed by strain-stiffening. 
          (B) Log-linear plots of the two relaxation times $\tau_{1}$ and $\tau_{2}$ show that they increase with applied strain and tend to saturate towards the higher values of strain (n=7).
          Error bars are SE.}
           \label{large-strain}  
           \end{center}
          \end{figure}

To overcome the limited travel range of the piezoelectric drive, which limits the strain range, we performed experiments by stretching silk using the motorized XYZ stage on to
which the piezo is mounted, and applied the superimposed strain oscillations using the piezo (see Material and Methods for details). The storage moduli obtained as before
shows a clear minimum at around 4--5\% strain when plotted as a function of pre-strain (Fig.~\ref{large-strain}(A), Fig. S7). The variation in the two relaxation times are shown in Fig.~\ref{large-strain}(B).
Both $\tau_1$ and $\tau_2$ increases with strain and tend to saturate towards higher pre-strain values. The maximum strain in these experiments was limited to about 30\% 
and we could observe signatures of partial breakage in some cases. 

\subsection*{Energy dissipation}

\begin{figure}[]
                            \begin{center}
                            \includegraphics[height=3in,width=6in]{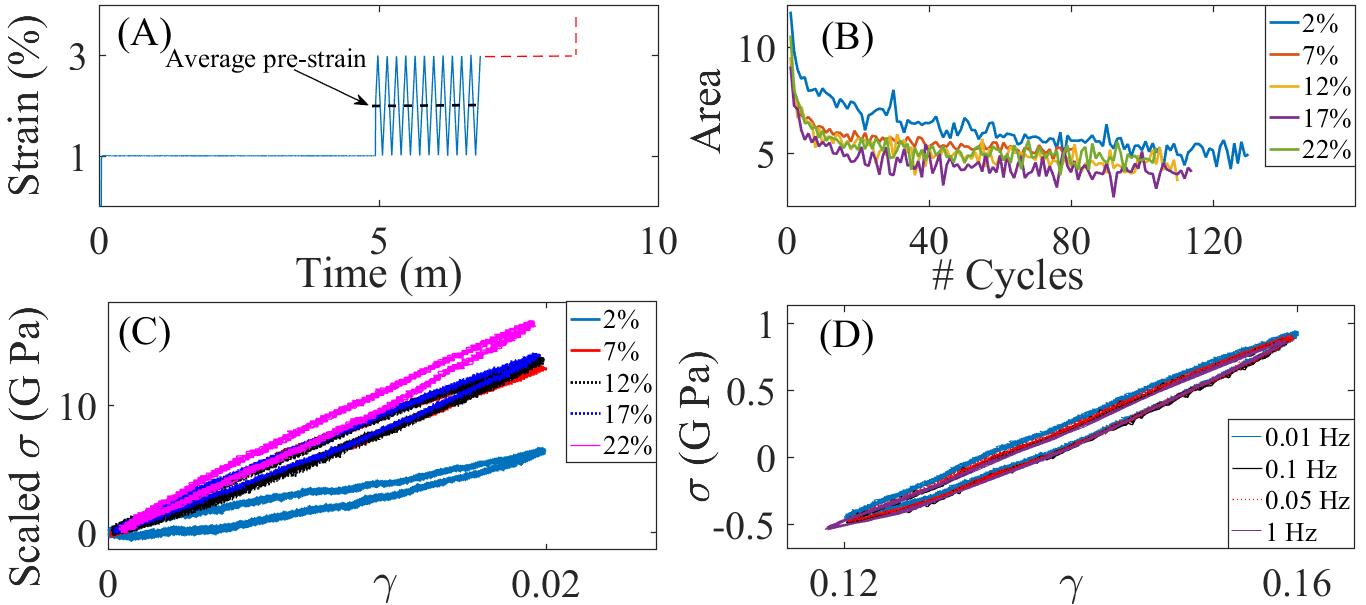}
                            \caption{(A) A schematic of the pre-strain plus cyclic ramp protocol used to obtain stress--strain  Lissajous figures in order to estimate the
                            dissipated energy. (B) The evolution of the enclosed area within each cycle with cycle number shows the trend towards a limit cycle.
                             (C) Lissajous figures obtained from limit cycles corresponding to various amounts of pre-strain. The pre-strain in each case is indicated in the figure and the strain amplitude is 2\% for all the cases. 
                            The data was rescaled by subtracting the lowest values of strain and stress 
                            ($\gamma(t)-\gamma_{\textrm{min}}$, $\sigma(t)-\sigma_{\textrm{min}}$) for representation.
                             (D) Experiments performed at different frequencies for a fixed pre-strain (strain amplitude 4\% ) show no significant change in the enclosed area of the limit cycle, within the frequency range allowed by the MER technique.}
                            \label{ramp-area}  
                            \end{center}
                            \end{figure}

The energy dissipated during cyclic deformations is proportional to the area enclosed by the stress vs. strain plot. 
In order to check if the energy dissipated during deformation has any strain dependence, we applied cyclic ramps (triangular perturbation instead of sinusoidal) as shown in Fig.~\ref{ramp-area}(A). The application of the triangular perturbation causes an increase of the average  pre-strain value. This leads to a relaxation of the average stress which finally reaches a steady-state (Fig. S8). Corresponding to this, the enclosed area for each cycle evolves until it too reaches a steady-state (at limit cycle) as shown in Fig.~\ref{ramp-area}(B). These limit cycles are used to study the dependence of dissipated energy on frequency or strain. In Fig.~\ref{ramp-area}(C), we show 
the Lissajous figures obtained from a few limit cycles for a ramp frequency of 0.05 Hz and amplitude of 2\% and for different amounts of average pre-strain. The increase in the values of peak stress with increasing average pre-strain value as shown in Fig.~\ref{ramp-area}(C) is due strain-stiffening of the silk fiber as discussed earlier. We also probed the energy dissipated as a function of frequency in the range 0.01--1 Hz and do not observe any variation within this range (Fig.~\ref{ramp-area}(D)).
               
 \subsection*{Ageing of silk}
\begin{figure}[!h]
         \begin{center}
       \includegraphics[width=6in]{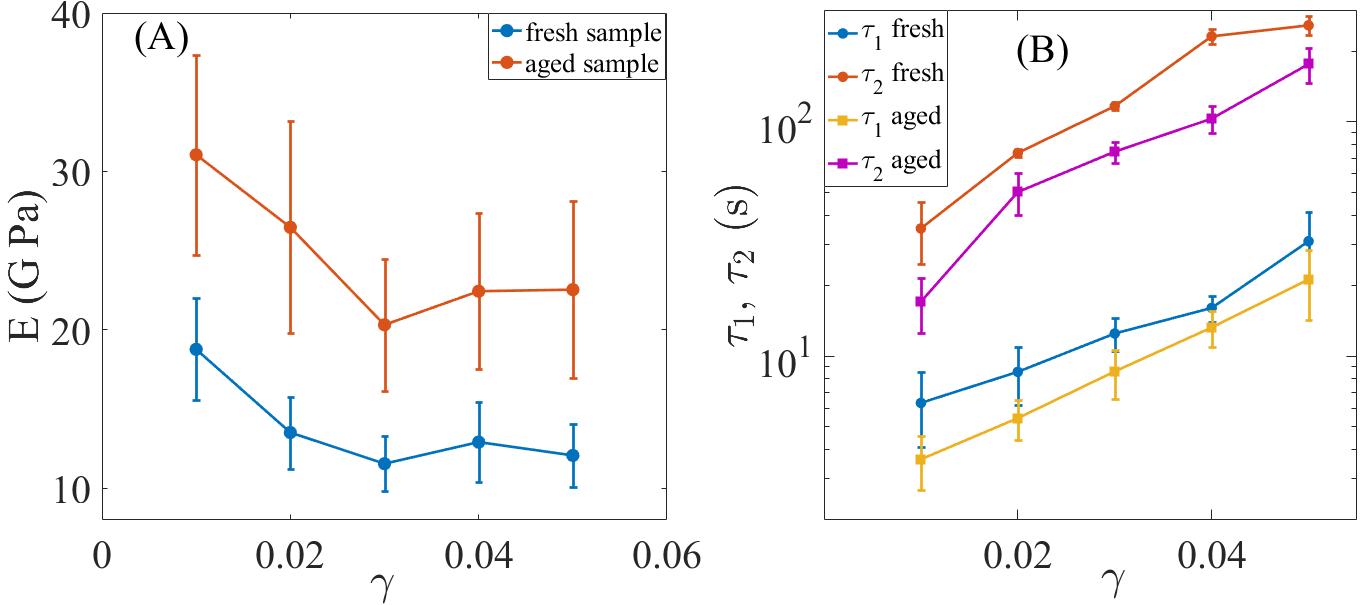}
        \caption{(A) Change in elastic moduli as a function of strain measured for fresh and aged silk strands. The same strand was measured a day after 
        extraction (fresh) and then 10--12 month later (aged). While the strain-softening response is retained, the moduli for the aged samples are significantly
        higher compared to measurements made soon after extraction. The error bars are SE (n=10).
        (B) The relaxation times obtained from fresh and aged silk fibers show similar increase with strain, but for a given pre-strain the relaxations become faster with age (n = 7). 
        }
         \label{age-effect}  
         \end{center}
        \end{figure}

Silk is expected to age due to effects like loss of water, degradation of the proteins, structural relaxation, etc. \cite{Lepore2016, Elices2005, Agnarsson2008}. 
In order to check whether the observed non-linear responses of silk are altered with age or not, we compared data obtained
 from the same fibers about day after spinning (``fresh'') and 10--12 month later (``aged''). The samples were stored in the lab as detailed in Material \& Methods.
 As can be seen from Fig.~\ref{age-effect}(A), the strain-softening behavior is retained in aged silk albeit with higher values of the storage moduli. The two relaxation times
 show the same trend with strain as fresh samples but the values, for a given strain, decrease with age (Fig.~\ref{age-effect}(B)), consistent with the increase in storage moduli.
 
 The ageing behaviour can be of interest in particular to social spiders as they show little tendency to abandon a web unless forced to by external agents (heavy wind or rain), since silk is expensive to produce. On the other hand, may solitary spiders will eat up their silk and go to another site \cite{Lubin2007, Peakall1971}. 
   
\section*{Discussion}

The results presented above may be summarised as follows. (i) We have probed the elastic response of social spider dragline silk fibers by applying a sequential step-strain plus small
amplitude sinusoidal oscillation protocol. This method  allows us to measure the elastic moduli as a function of strain under a quasi-equilibrium condition. This is unlike previous
reports where, typically, the silk strand is extended continuously to highly non-linear regimes at a given strain rate. (ii) The measurement of the elastic moduli at different
pre-strains clearly show an initial strain-softening response which lasts till about 4\% strain followed by a strain-stiffening response. (iii) By Fourier decomposition of the stress
relaxation subsequent to each step-strain, we are able to extract the frequency response of the storage and loss moduli for different values of pre-strain. (iv) We also show that the
stress relaxation curves can be fitted well to a double exponential function. The two relaxation times thus obtained monotonically increase with pre-strain whereas the elastic moduli 
goes through a minimum at around 4\% strain. We expect these novel results to place additional constraints on the various models for spider dragline silk fibers.
Below we present a short discussion of current models and discuss our results in this context.

Dragline silk proteins Spidroin I and Spidroin II consist of alanine rich anti-parallel beta-sheets organised as nano-crystallites which are a few nano-meters in size, and glycine rich amorphous regions which are connected together via hydrogen bonds \cite{Sponner2005, Blamires2017}. Single fiber X-ray diffraction studies show that the crystalline regions are oriented with the beta-sheets parallel to the fiber axis \cite{Glisovic2008}. Recent NMR studies indicate that the amorphous regions are also orientationally ordered with the helical secondary structure aligned along the fiber direction \cite{Beek2002}. The degree of order in the crystalline and amorphous regions as well as their relative fractions may vary between spider species, and also with humidity \cite{liu2005, yarger2018, Blamires2017, Liu2005b}. Application of strain results in a reduction in the size of the crystallites, suggesting unfolding of the beta-sheets under force \cite{Glisovic2008}. Although there are several studies on the microstructure of silk using Raman scattering, NMR, and X-ray, the relation between the microstructure and mechanical response is still not fully understood \cite{Glisovic2008, Simmons1996, Creager2010, Lefvre2012, Shao1999b}.

Various models have been proposed to account for the force-extension curves of dragline silk fibers. Termonia \cite{Termonia1994} considered breaking of hydrogen bonds within the amorphous regions as an activated process with a force dependence to arrive at force-extension plots for wet and dry dragline fibers, similar to those seen in experiments. The polymer chains devoid of hydrogen bonds are modelled as a non-linear elastomer and the nano-crystals are assumed to be rigid without any force dependent structural changes. Such a model qualitatively reproduces the experimentally
observed responses of dry silk and hydrated silk where hydrogen bonds are assumed to be in dissociated state. In the case of dry silk one observes the characteristic kink in the force extension curve which is absent in the case of hydrated silk fibers.  

Tommasi et al., \cite{Tommasi2010} developed a model where the silk fiber is considered to be a composite material consisting of crystalline and amorphous regions, as described by earlier models. The stiff crystalline regions are assumed to be a linear elastic material which can undergo a transition to amorphous beyond a threshold strain. This is motivated by AFM experiments on single silk proteins which show force induced unfolding of domains \cite{Oroudjev2002}. The amorphous regions are treated as Worm-Like-Chain (WLC) entropic springs with characteristic non-linear strain-stiffening response. By changing the relative amounts of these two regions, pure stiffening (high amorphous content) or softening followed by stiffening (intermediate amorphous content) responses can be obtained. The initial softening is due to the unfolding events which dominate at small strains, above a force threshold, and the subsequent stiffening may be attributed to the non-linearity of the WLC elasticity.

While these models are able to reproduce some of the main qualitative behaviours of silk fibers, and provide some insight into the microscopic processes occurring in dragline
fibers under extension, they lack the ability to predict time-dependent properties like the stress relaxation or the frequency dependence of the elastic and loss moduli presented in this article.
Thus, the experiments presented here should motivate models that incorporate the kinetics of unfolding or bond dissociation explicitly which can then yield time-dependent
properties. An attempt of this nature has been successfully employed recently to describe the non-linear elasticity and relaxation behaviour of axons, which are tubular extensions
of neuronal cells \cite{Dubey2019}. We hope that the results we present here will motivate the development of such models to better understand the origins of the fascinating properties of spider dragline silk.

\section*{Acknowledgements}
We thank Tejas G Murthy for initial discussions, Thejasvi Beleyur for help with the initial trials and Yatheendran K M for performing the electron microscopy. SM thanks SERB (under DST, Govt. of India) for support through a Ramanujan Fellowship. Authors declare no conflict of interest.
    
\newpage

\bibliographystyle{unsrtnat}

\bibliography{References}


%
%
%
%
%
%
%
%
%
%
%
%
%

%
%
%


\renewcommand{\figurename}{Fig.}
\renewcommand{\thefigure}{S\arabic{figure}}
\renewcommand{\thetable}{S\arabic{table}}


%
%
%

\setcounter{figure}{0}    

\section*{SUPPLEMENTARY  MATERIALS:}
\section{Extraction of Silk for force measurement}

\begin{figure}[!h]

\begin{center}  
\includegraphics[width=6in]{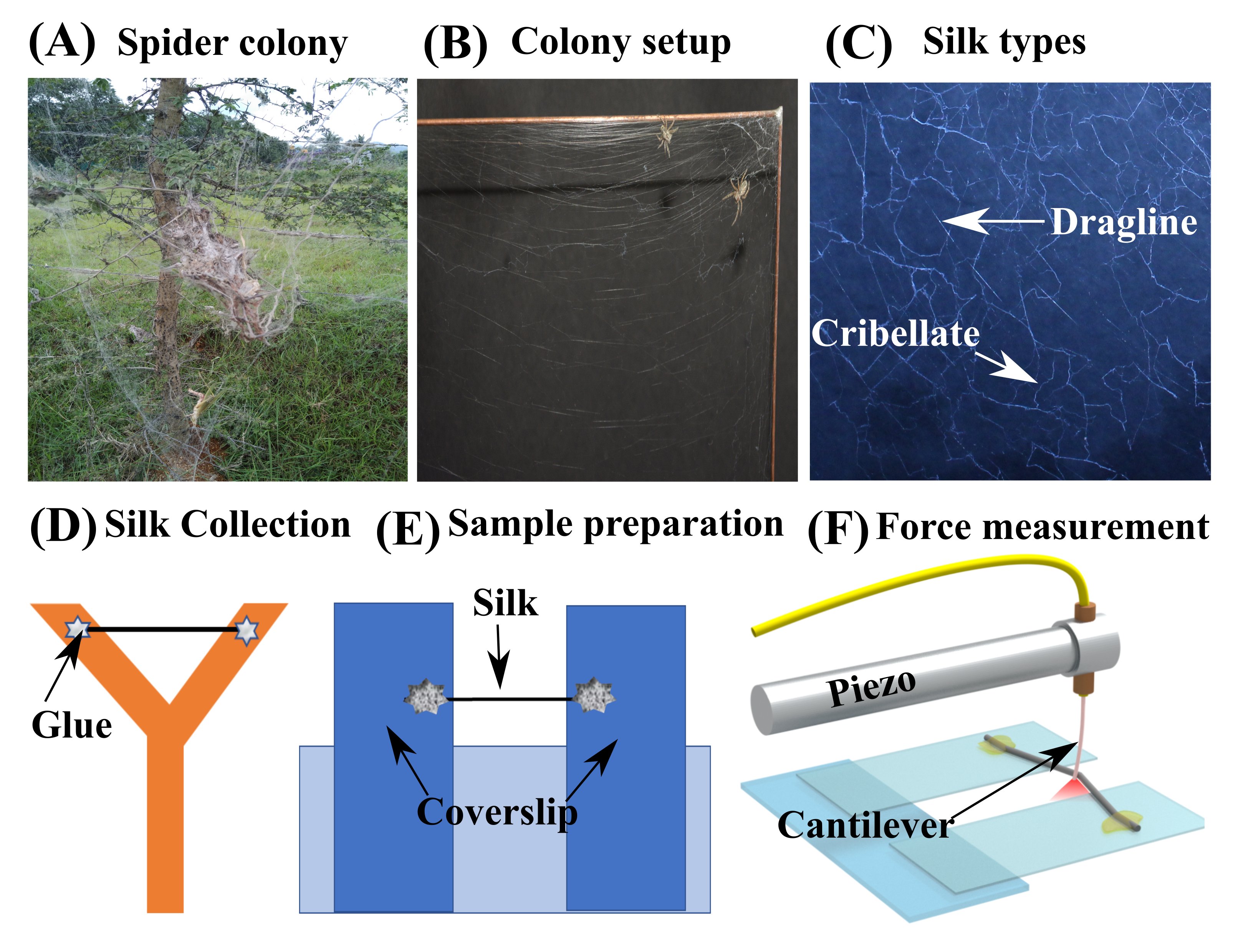}  
\caption{(A) Photograph of a web of social spider {\it Stegodyphus sarasinorum} in the wild. (B) Photograph of spiders with their web built on a metal wire frame
as detailed in the Material and Methods. (C) Image showing an expanded view of the web with dragline 
and cribellate silk indicated by the labels. (D) Schematic of a silk strand transferred to a Y-shaped cardboard frame. (E) Schematic of the silk strand after it has been
transferred  to a frame made of two coverslips attached to a glass slide. (F) Schematic of the a silk fiber stretched using an optical fiber cantilever attached to piezo.
}
\label{}  
\end{center}  
\end{figure}

\newpage
\vspace{-1cm}
\section{Images of silk being pulled using optical fiber}
\begin{figure}[!h]
\begin{center}  
\includegraphics[width=4in]{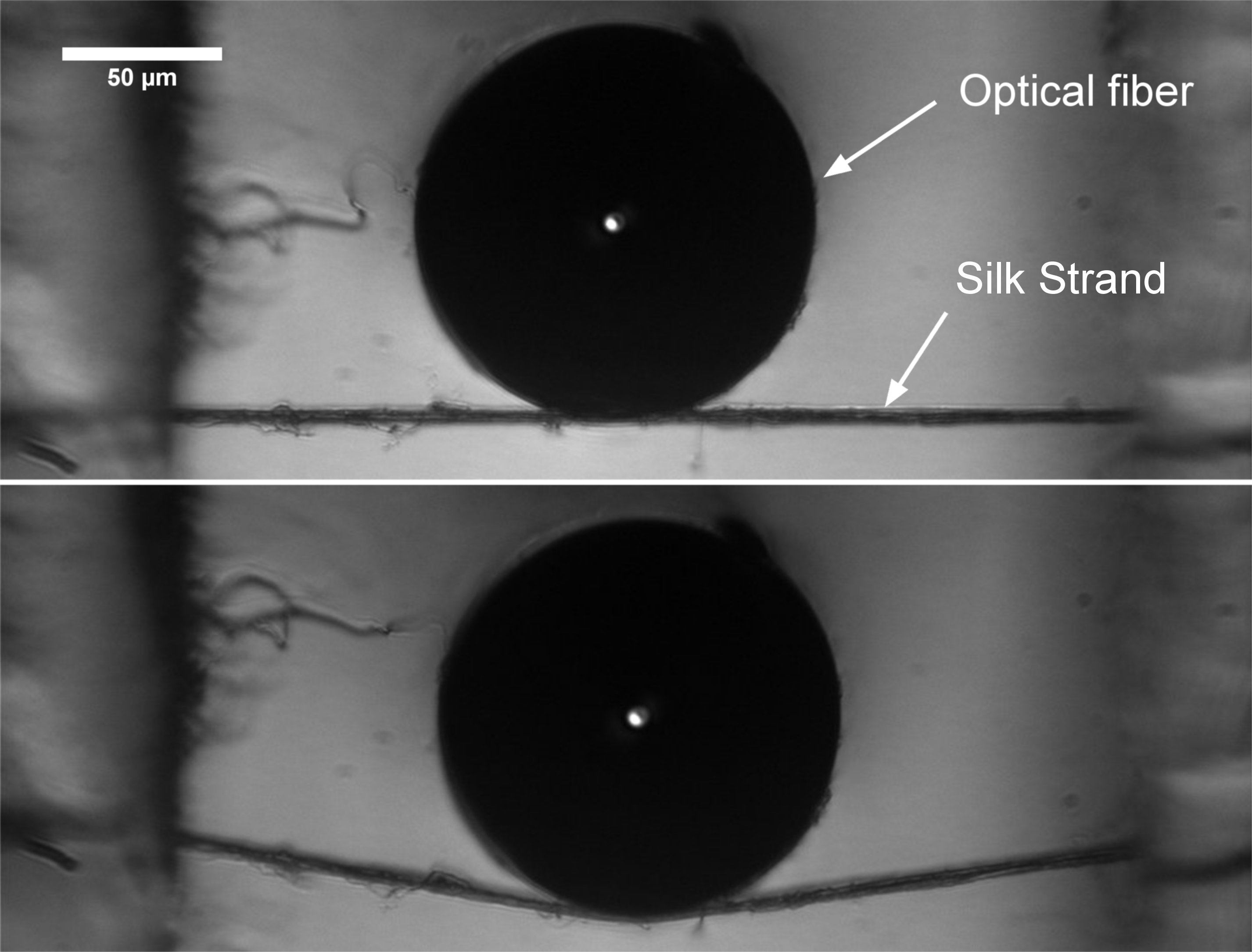}  
\caption{A typical images of a silk strand which is stretched using an optical fiber in transverse direction.
Images of silk strand before pulled (top) and after pulled (bottom). The bright spot is laser light exiting the core of the optical fiber which is detected using a Position Sensitive Detector (PSD).}
\label{}  
\end{center}  
\end{figure}

\newpage

\section{A Photograph of the Micro-Extension Rheometer}
\begin{figure}[!h]
\begin{center}  
\includegraphics[width=5in]{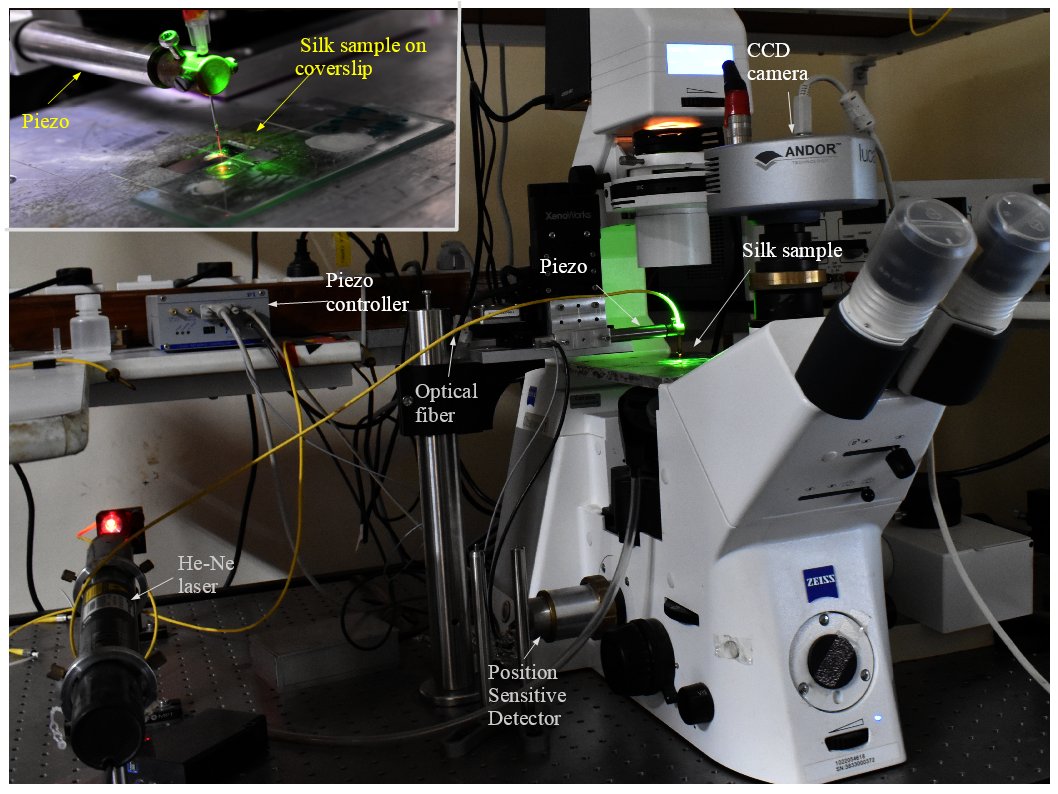}  
\caption{Photograph of the optical fiber based force apparatus Micro-Extension Rheometer (MER) used to investigate the
mechanical properties of spider silk. The piezo drive and the Position Sensitive Detector (PSD) are interfaced to
a computer to operate the setup in a feedback strain control mode. More details are described in the main text. The inset shows the close up view of the sample which is being pulled with the optical fiber.}
\label{}  
\end{center}  
\end{figure}

\newpage
\section{Electron microscopy images of dragline and cribellate silk}
\vspace{1in}
\begin{figure}[!h]
\begin{center}  
\includegraphics[width=6in]{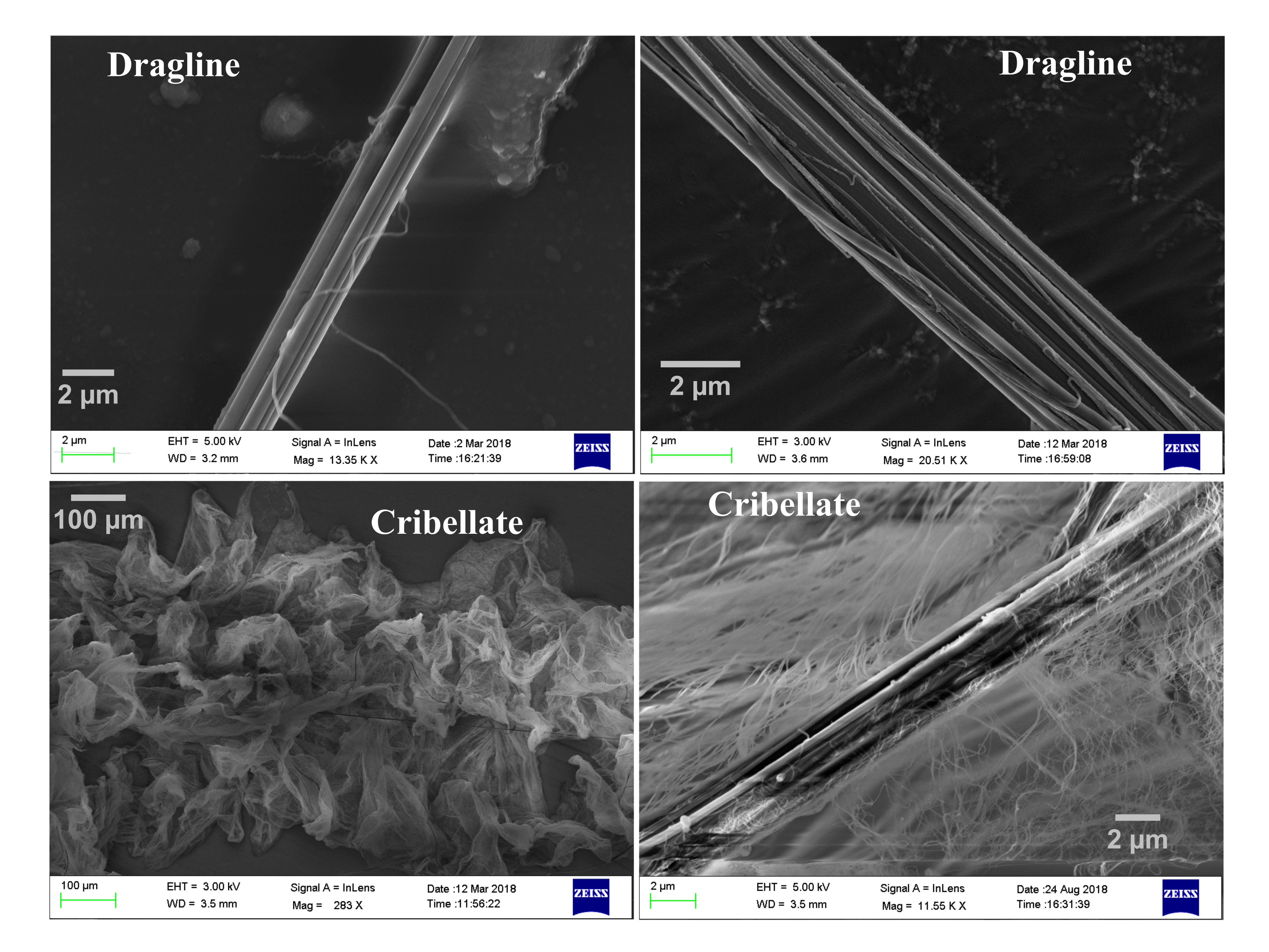}  
\caption{
Electron Microscopy images of dragline and cribellate silk: Dragline is made of many thin fibers whereas cribellate consist of a central fiber and surrounded by ultrathin fibers which are sticky in nature. 
}
\label{}  
\end{center}  
\end{figure}

        
\newpage
\section{Pre-Tension for different silk strands}

\begin{figure}[!h]
\begin{center}  
\includegraphics[width=6in]{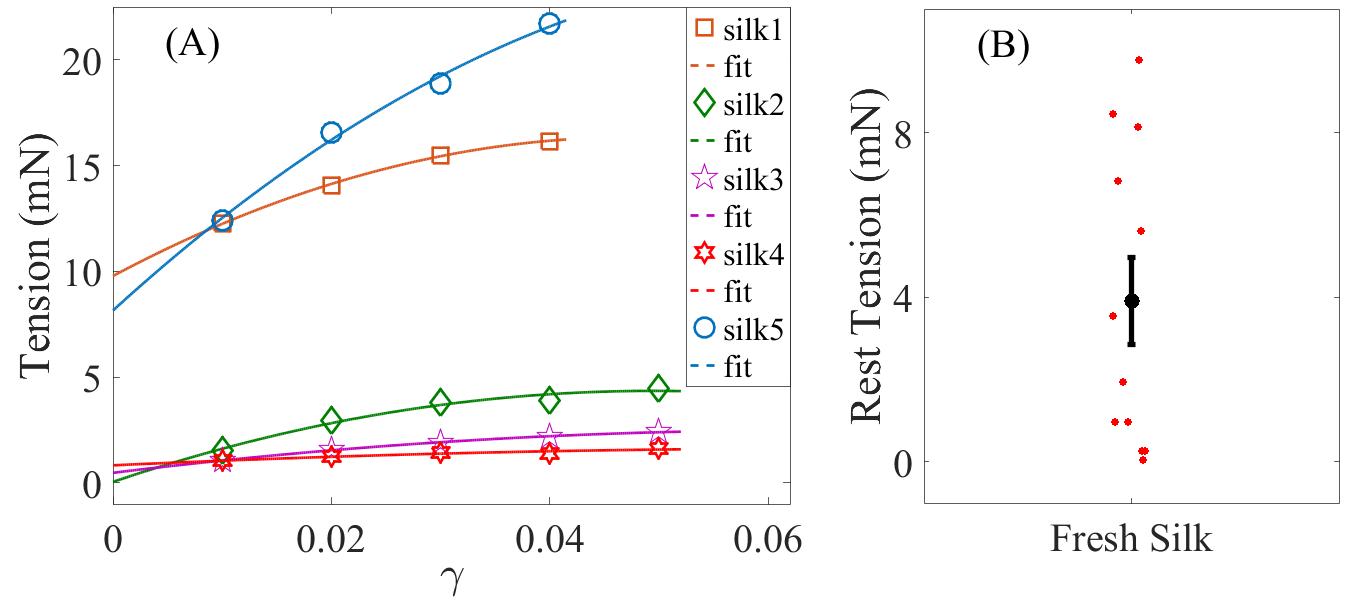}  
\caption{The silk strands are under pre-tension when they are in the web. (A) This pre-tension is obtained by extrapolating the steady state tension vs. strain to zero strain by fitting the data with the equation $(-ax^2+bx+c)$. Fits for different silk strands are shown. (B) Pre-tension values for different silk strands (red dots) (The black dot and the error bar are the mean and standard error, respectively.). }
\label{}  
\end{center}  
\end{figure}

\newpage
\section{Fourier Transform Analysis}

\begin{figure}[!h]
\begin{center}  
\includegraphics[width=4in]{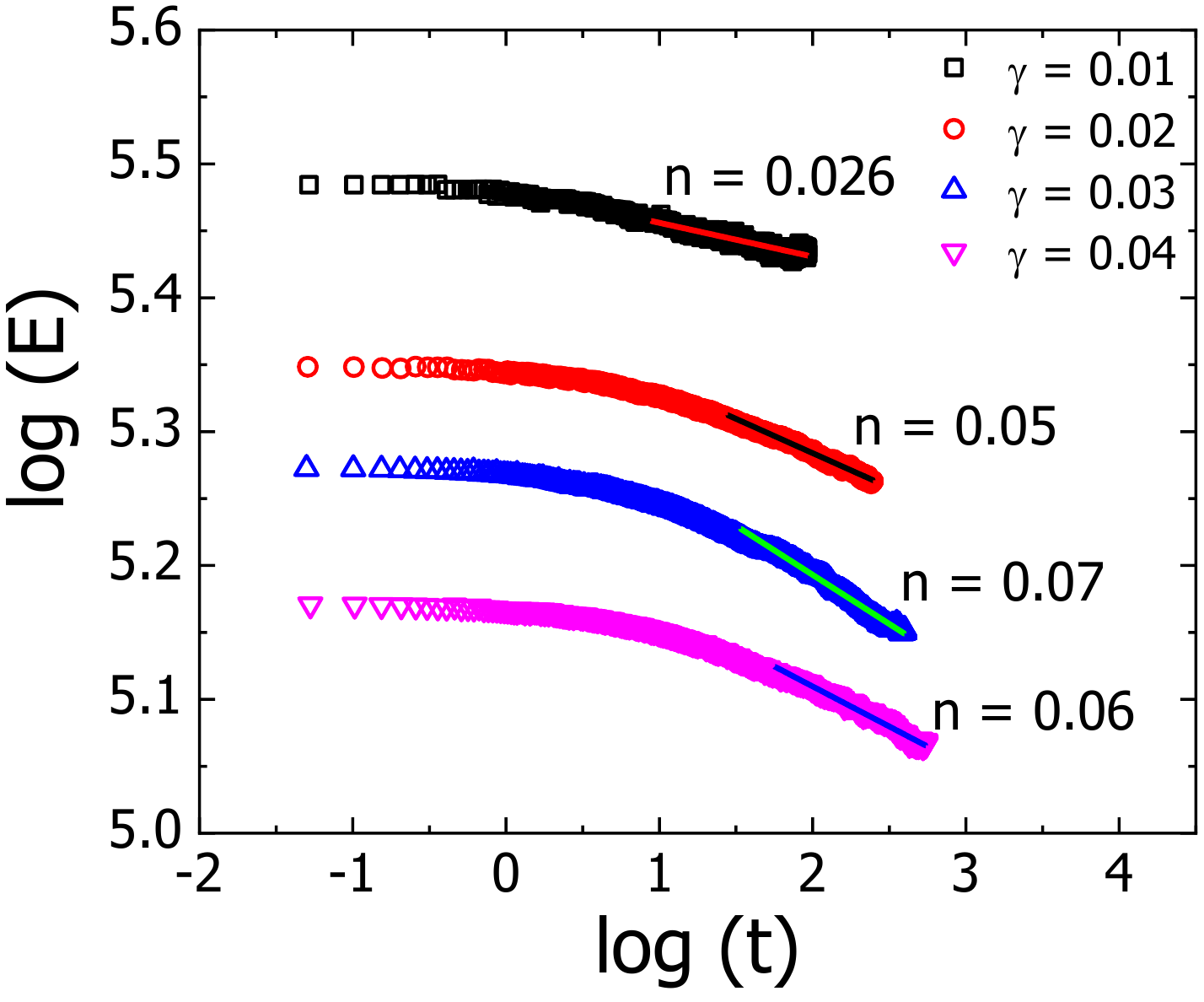}  
\caption{Variation of stress relaxation modulus $E$ (defined in the main text)as a function of time. For convenience, logarithm of the quantities are shown. The magnitudes of the logarithmic slope $n$ (described in the main text) are indicated in the figure for different applied step-strain values.}
\label{}  
\end{center}  
\end{figure}

\newpage
 \section{Normalised modulus in large strain regime}
    
    \begin{figure}[h!]
         \begin{center}
       \includegraphics[width=5in]{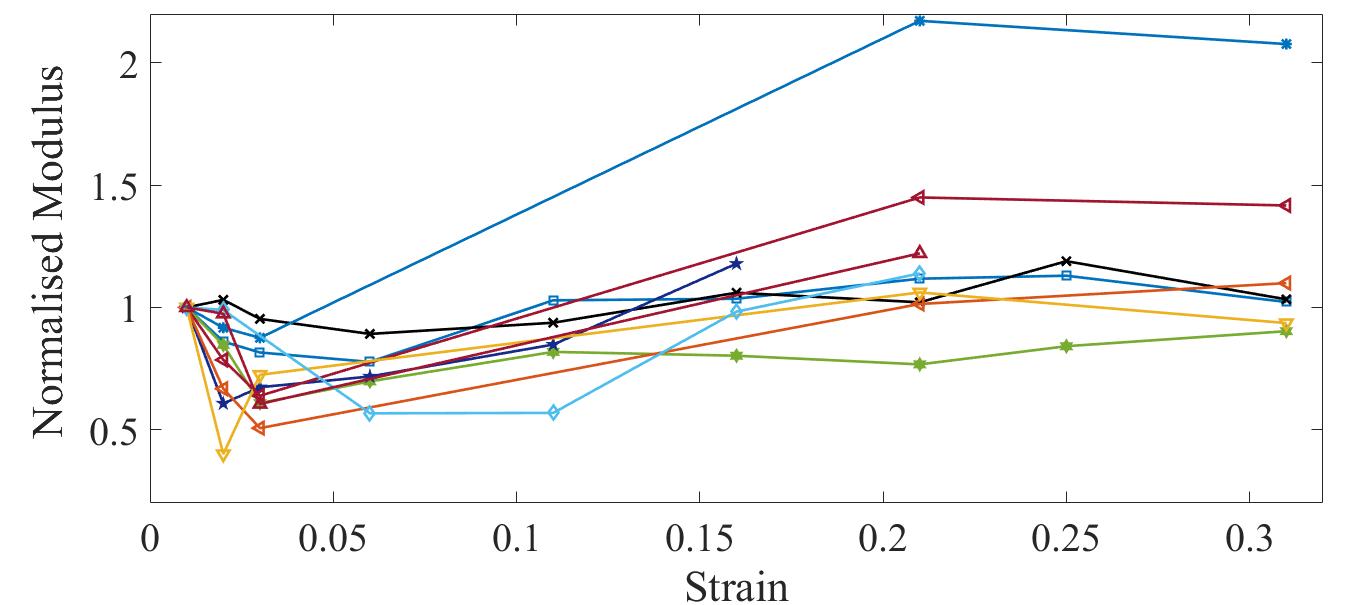}
        \caption{Elastic modulus normalised with respect to its initial value shows that there is an initial softening.}
         \label{}  
         \end{center}
        \end{figure}
     
\section{Ramp Exp}
                             
  \begin{figure}[h!]
     \begin{center}
   \includegraphics[width=5in]{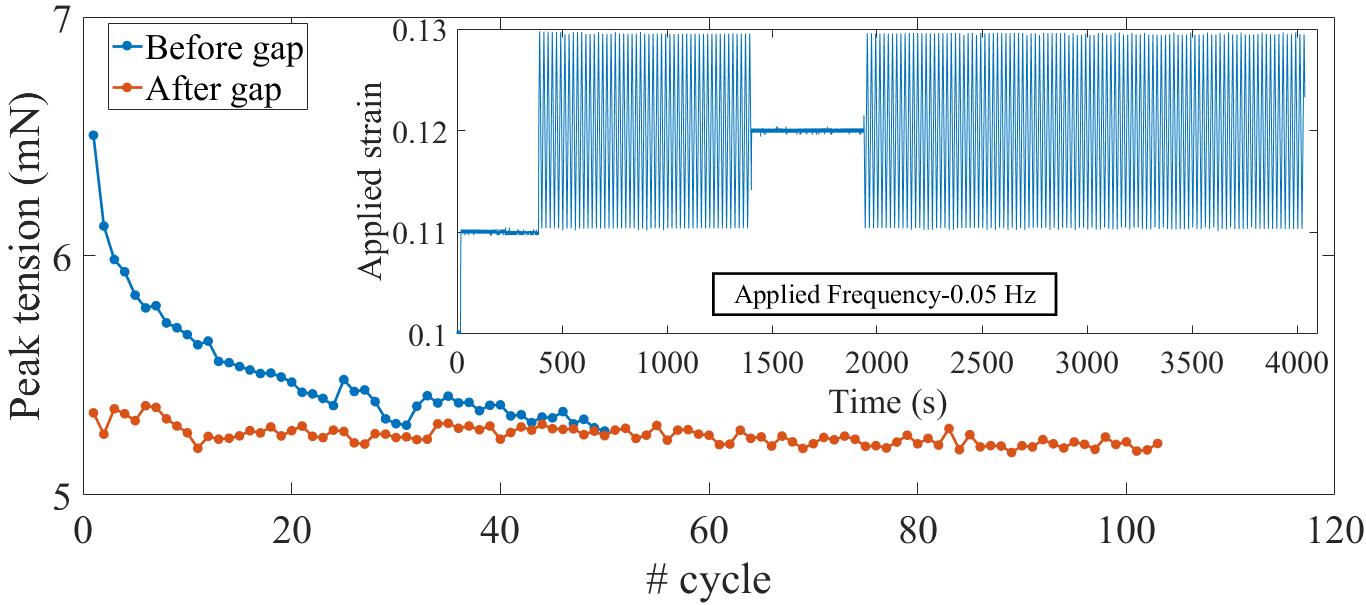}
    \caption{Variation in peak tension with cycle number  obtained from superimposed triangular wave soon after a strain step (blue dots) and that obtained after a wait time without superimposed triangular wave (orange dots). This data shows that the silk fiber reaches a limit cycle after approximately 50 cycles.}
     \label{}  
     \end{center}
    \end{figure}



%
%
%
%

\end{document}